\documentclass[preprint,12pt]{elsarticle}
\usepackage{graphics}
\usepackage{graphicx}
\usepackage{epsfig}
\usepackage{subfigure}
\usepackage{amssymb}
\usepackage[dvips]{color}
\journal{Solid State Sciences}
\begin{document}
\begin{frontmatter}
\title{Structure, elastic and dynamical properties of KN$_3$ and RbN$_3$: A van der Waals density functional study}
\author{K. Ramesh Babu and G. Vaitheeswaran}
\address{Advanced Centre of Research in High Energy Materials (ACRHEM),
University of Hyderabad, Prof. C. R. Rao Road, Gachibowli, Andhra Pradesh, Hyderabad - 500 046, India\\
\vspace{0.6in}
*Corresponding author E-mail address: gvsp@uohyd.ernet.in\\
\hspace {1.2in} Tel No.: +91-40-23138709\\
\hspace {1.2in} Fax No.: +91-40-23010227}
\end{frontmatter}
\clearpage
\section*{Abstract}
We report a detailed first principles study on the structural, elastic, vibrational and thermodynamic properties of layered structure energetic alkali metal azides KN$_3$ and RbN$_3$. All the calculations were carried out by means of plane wave pseudopotential method with and without including van der Waals interactions. The calculated ground state structural properties are improved to a greater extent by the inclusion of dispersion corrections, implies that the van der Waals interactions play a major role on the physical properties of these systems. The elastic constants and the related bulk mechanical properties for the tetragonal KN$_3$ and RbN$_3$ have been calculated using both the methods and found that the compounds are mechanically stable systems. The magnitude of the calculated elastic constants increases in the order RbN$_3$ $<$ KN$_3$ implying higher elastic stiffness for KN$_3$, the fact also confirmed by the higher values of bulk, shear and Young's moduli of KN$_3$ than RbN$_3$. Moreover, the calculated elastic constants follows the inequality C$_{33}$ $<$ C$_{11}$ which indicates the presence of more number of intermolecular interactions along a-axis over c-axis of the azide lattices. A correlation has been proposed to relate the calculated elastic constants to the decomposition phenomena for the metal azides. The experimentally reported vibrational frequencies at the gamma point were exactly reproduced by the present calculations. In addition, we also present the thermodynamic properties such as heat capacity which compares well with the experiment.\\
Keywords:
van der Waals interactions; elastic constants; vibrational frequencies; thermodynamic properties
\newpage
\section{Introduction}
\label{}
 Energetic materials have important applications in defense, space and civilian sectors. Energetic materials release huge amount of heat energy accompanied with large amount of gaseous products upon external stimulation such as heat, shock, light. Inorganic metal azides can be used to initiate the well known high energetic materials such as RDX, HMX, and CL-20 etc \cite{Akhavan}. Among the inorganic metal azides, alkali metal azides are structurally simple, relatively stable and are used as starting materials in synthesis of polymeric nitrogen, an ultimate green high energy material known till so far \cite{Eremets}. The structural simplicity of these compounds enable us to understand the complex chemical phenomena such as decomposition and detonation that occur in solid state energetic materials \cite {Fair, Bowden}. Alkali metal azides, similar to other metal azides, show variation in stability towards heat, light, and shock. They undergo decomposition into metal atom and nitrogen gas under the action of suitable wave length of light (ultra-violet light) and even detonation may occur when the temperature is close to their melting point. The decomposition reaction is as follows:
 \begin{equation}
 2MN_3\rightarrow 2M + 3N_2
 \end{equation}
 where `M' is the metal atom. Since the final decomposition product is pure nitrogen gas, these materials can also be used as gas generators and also in automobile air bags \cite{Evans}.
\paragraph*{} Alkali metal azides, KN$_3$ and RbN$_3$ crystallize in a body-centered tetragonal structure with space group I4/mcm (140). The unit cell contains four molecules per unit cell. The crystal structure of the metal azides clearly shows that the arrangements of the atoms are in layers as the azide ions are sandwiched between the metal atom layers \cite{Muller}. This structural anatomy leads to a strong bonding between these two layers whereas there also exists a non-bonded interactions between the non-bonded nitrogen atoms of the azide ion also called as van der Waals interactions as shown in Fig 1. Thus, the crystal structure of the metal azides involve both bonded and non-bonded interatomic interactions. Recently, the high-pressure effect on these alkali metal azides has become a promising topic because of their application as precursors to synthesize a high energetic density material of a polymeric form of nitrogen \cite{Chen, Medvedev-2, Ramesh}.
\paragraph*{}Both KN$_3$ and RbN$_3$ undergo rapid decomposition with evolution of large amounts of pure nitrogen gas at temperatures close to their melting points. The metal azides show variation in impact sensitivity towards mechanical shock. KN$_3$ is not sensitive to impact or friction whereas RbN$_3$ is more sensitive to mechanical shock and it compares with that of impact sensitivity of trinitrotoluene (TNT) \cite{Fair}. It is a known fact that decomposition and detonation are the collective properties of materials which depends on intermolecular interactions, molecular arrangements and thereby the crystal structure of the solid state. Thus, it is crucial to know the basic crystal structure and also the structure-related properties such as elastic and vibrational properties for better understanding of the nitrogen release decomposition phenomena of the metal azides and their variation in sensitivity towards mechanical shock. Also, these properties are important to understand the high pressure behavior of the metal azides.
\paragraph*{} Density functional theory (DFT) is the robust theoretical approach to predict the material properties. But, the conventional DFT techniques based on local density approximation (LDA) and generalized gradient approximation (GGA) will tend to give large errors when compared to experiments due to the lack of description of vdW interactions \cite{Jones}. However, there are very less number of studies available on these metal azides based on DFT. These studies were mainly devoted to electronic structure \cite{Gord, Ju} and optical properties \cite{Yu, Zhu}. In our earlier work, we have also studied the structural and electronic properties of LiN$_3$ and KN$_3$ \cite{Ram} as well as the vibrational properties of LiN$_3$ and KN$_3$ by using DFT and dispersion corrected DFT calculations \cite{Ramesh, Ramesh-1}. To the best of our knowledge, there are no reports available on the comparative study of structure related properties such as elastic constants, phonon properties and thermodynamic properties of the metal azides KN$_3$ and RbN$_3$ based on density functional calculations with and without accounting the vdW interactions.
In this work, efforts have been taken to study the effect of vdW interactions on ground state structural, elastic, vibrational and thermodynamic properties of both the metal azides KN$_3$ and RbN$_3$. We have also studied the electronic band structures and nature of bonded interactions from density of states and also through bond population analysis. The calculated elastic constants were used to understand the sensitivity of the metal azide crystal lattice to the external stress. The computed vibrational frequencies and phonon density of states were used to calculate the heat capacity of the azides.
\paragraph*{} The rest of the paper is organized in the following manner. The second section describes the computational details used for the present study. The results and discussion are presented in section 3. Finally in section 4 we end with a brief conclusion of our study.
\section{Computational details}
\label{}
First-principles density functional theory calculations were performed with the Cambridge Sequential Total Energy Package (CASTEP) program \cite{Payne, Segall}, using Vanderbilt-type ultrasoft pseudo potentials \cite{Vanderbilt} and a plane wave expansion of the wave functions. The electronic wave functions were obtained using  density mixing scheme \cite{Kresse} and the structures were relaxed using the Broyden, Fletcher, Goldfarb, and Shannon (BFGS) method \cite{Fischer}. The exchange-correlation potential of Ceperley and Alder \cite{Ceperley} parameterized by Perdew and Zunger (CA-PZ) \cite{PPerdew} in the local density approximation (LDA) and also the generalized gradient approximation (GGA) in Perdew-Burke-Ernzerhof (PBE) \cite{Perdew} was used to describe the exchange-correlation potential. The pseudo atomic calculations were performed for K 3$s^2$ 3$p^6$ 4$s^1$, Rb 4$s^2$ 4$p^6$ 5$s^1$,  Cs 5$s^2$ 5$p^6$ 6$s^1$ and N 2$s^2$ 2$p^3$. The Monkhorst-Pack scheme k-point sampling was used for integration over the Brillouin zone \cite{Monkhorst}. It is well known that the cut-off energy and k-point mesh influences the convergence of calculations. Hence we tested the dependence of energy cut-off and k-point grid and found that for 520 eV plane wave cut-off energy and 5x8x5 k-point mesh, the change in total energy is less than 1meV. We have then chosen these plane wave cut-off energy and k-point mesh for all the calculations.
 In the geometry relaxation, the self-consistent convergence on the total energy is 5x10$^{-7}$ eV/atom and the maximum force on the atom is found to be 10$^{-4}$ eV/{\AA}. Elastic constants were calculated using finite strain technique as implemented in CASTEP code. We have used primitive cell that contains two molecules for the calculation of vibrational frequencies. All the dynamical calculations were carried out by using the density functional perturbation theory as implemented in the CASTEP code \cite{Gonze, Refson}.
\subsection{Dispersion corrections to the DFT}
 To treat vdW interactions efficiently, we have used the vdW correction to the exchange - correlation functional of standard density functional theory at semi empirical level. According to semi-empirical dispersion correction approach, the total energy of the system can be expressed as
\begin{equation}
E_{total} = E_{DFT} + E_{Disp}
\end{equation}
where
\begin{equation}
E_{Disp} = s_i\Sigma_{i=1}^N\Sigma_{j>i}^Nf(S_RR^{0}_{ij}, R_{ij})C_{6, ij}R_{ij}^{-6}
\end{equation}
here C$_{6, ij}$ is called dispersion coefficient between any atom pair $i$ and $j$ which solely depends upon the material and R$_{ij}$ is the distance between the atoms $i$ and $j$ respectively. In the present study we have used the recently developed schemes by Ortmann, Bechstedt, and Schmidt (OBS) \cite{OBS} in LDA and Grimme (G06) \cite{Grimme}, Tkatchenko and Scheffler (TS) \cite{TS} approaches in GGA. These semiempirical approaches provide the best compromise between the cost of first principles evaluation of the dispersion terms and the need to improve non-bonding interactions in the standard DFT description.
\section{Results and discussion}
\subsection{Crystal structure}
As a first step we computed the equilibrium crystal geometries of the two azides KN$_3$ and RbN$_3$ by allowing the full lattice and atoms to relax and then computed the total energies. The structures with minimum total energy were considered as the optimized crystal structures and are used for further calculations. Although these materials are analogous to simple alkali halides but the presence of linear azide ion makes the crystal structure complex with weak dispersion interactions between the azide ion layers. In our previous work, we applied three different functionals (LDA (CA-PZ), GGA (PBE) and GGA (PBE+G06)) to potassium azide, KN$_3$, \cite{Ramesh, Ram} as a test and found that the GGA (PBE+G06) functional could reproduce the crystal structure that is in close agreement with the experimental structure. In continuation to our earlier work in this present study, we have used the recently developed dispersion corrected functionals LDA (OBS), GGA (PBE+TS) in addition to the LDA (CA-PZ), GGA (PBE) and GGA (PBE+G06) functionals in order to check their virtue in describing the crystal structure of KN$_3$ and RbN$_3$. The calculated values of the lattice parameters of the two azides within the density functionals of LDA (CA-PZ), GGA (PBE) and dispersion corrected functionals LDA (OBS), GGA (PBE+TS) and GGA (PBE+G06) along with experimental data \cite{Muller} are shown in Table 1. When compared to the experimental data our LDA calculations underestimate the lattice volume by -12.7$\%$ for KN$_3$ and -10.6$\%$ for RbN$_3$ whereas our GGA calculations overestimate the same by 4.6$\%$ for KN$_3$ and 6.7$\%$ for RbN$_3$ as shown in Table 1. This discrepancy can be expected for general density functional theory calculations using LDA and GGA \cite{Jones}. However, the calculated volume of the azides using the dispersion corrected functionals are underestimated for both the compounds. Among the three dispersion corrected functionals, PBE+G06 gives the theoretical equilibrium volume that is very close to the experimental results (see Table 1). The theoretical volume obtained using PBE+G06 functional results with an underestimation of -2.2$\%$ for KN$_3$ and -0.04$\%$ for RbN$_3$ respectively. Whereas the LDA (OBS), GGA (PBE+TS) functionals give the theoretical results that are smaller than the measured values by -17.9$\%$, -12.9$\%$ for KN$_3$ and -14.8$\%$, -16.3$\%$ for RbN$_3$ respectively. Therefore among the three dispersion corrected functionals, the crystal volume calculated with PBE+G06 functional gives best agreement with that of the experimental volume.
\paragraph*{}The predicted lattice parameter `a' which involves the non-bonded interactions i.e., vdW forces between the layers, see Fig. 1, using the conventional LDA (CA-PZ) and GGA (PBE) functionals result in large errors when compared to the experimental values by -3.5$\%$, 1.5$\%$ for KN$_3$ and -3.1$\%$, 2.1$\%$ for RbN$_3$ respectively. The predicted lattice constant `a' using the dispersion corrected functionals show large variations in terms of functionals where the LDA (OBS) and GGA (PBE+TS) underestimate by 4.8$\%$, 8.1$\%$ for KN$_3$ and also by 4.2$\%$, 8.7$\%$ for RbN$_3$ respectively. This readily tells that these functionals are not suitable for the description of the crystal structures of the metal azides. In comparison with experimental results, the functional PBE+G06 underestimates the value of `a' by -0.1$\%$ for KN$_3$ and overestimates by 0.7$\%$ for RbN$_3$. However, the results obtained by the dispersion corrected functionals are consistent with the trend reported in previous literature on molecular solids \cite{Bheem} and layered compounds such as black phosphorus \cite{Appa}. For the case of lattice parameter `c', the functionals followed their general trends with LDA (OBS) results in large errors of about -9.3$\%$ for KN$_3$ and -7.1$\%$ for RbN$_3$. Both GGA (PBE) and GGA (PBE+G06) predicted the lattice constant `c' with error of about 1.5$\%$, -1.8$\%$ for KN$_3$ and 2.2$\%$, -1.4$\%$ for RbN$_3$, respectively. Overall, the best qualitative agreements between the experimental results and theory for the lattice constants are achieved by the PBE+G06 functional. Therefore from the present study of crystal structures of the metal azides using various functionals we propose that PBE+G06 functional give best results and it can be employed for the structural description of the other metal azides for future studies.
\paragraph*{} It is very well known that the usual LDA calculations over estimate and GGA calculations under estimate the bulk modulus (B) by 10-12 $\%$ \cite{Jones}. The present calculations also follow this usual trend for both the compounds. Whereas, the bulk modulus calculated using the dispersion corrected functionals generally overestimate because of the low equilibrium volume. Among the three dispersion corrected functionals, PBE+G06 gives bulk modulus value that is in good agreement with experiment. The bulk modulus value of KN$_3$ calculated using PBE+G06 is 27.2 GPa, which is in good agreement with the experimental value of 27.4 GPa \cite{Medvedev-2}. It should be noted here that, in our earlier work on KN$_3$, the bulk modulus was calculated to be 20.4 GPa using PBE functional and 26.3 GPa by using the PBE+G06 functional \cite{Ramesh}. The present calculated value of bulk modulus differs slightly with these reported values. This is because the present value of `B' is calculated from the elastic constants whereas the reported values are obtained from the fitting of pressure-volume data to equation of state. The disagreement arises from a variety of sources, both numerical and theoretical, as the two methods are very different in detail. In the EOS approach, a series of volumes is chosen at optimization of different pressures. The pressure-volume data is then fitted to Equation of State (Murnaghan equation of state) and the bulk modulus extracted from the fit. On the other hand, calculating the bulk modulus from the elastic constants involves a series of displacements along the crystalline axes, with optimization of internal co-ordinates at each displacement, using analytic first-derivatives and numerical second-derivatives of the total energy with respect to displacement taken, resulting in the elastic constants, which are then used to find the compliance matrix elements and the bulk modulus. Therefore one would expect a slight variation in the obtained values of the bulk modulus. The bulk modulus value of RbN$_3$ is obtained to be 25.2 GPa. There is no experimental report available to compare the present `B' value of RbN$_3$. However, the `B' value of RbN$_3$ is lower than that of KN$_3$ because of the high equilibrium volume of the initial crystal geometry. One should note that the bulk modulus value of KN$_3$ and RbN$_3$ is larger than that of corresponding metals K (3.4 GPa) and Rb (2.7 GPa) \cite{Gabor}. In general nitride materials have bulk modulus larger than 300 GPa which enables them as hard and ultra-incompressible materials \cite{Errandonea}. Whereas, the present metal azides with vdW interactions are highly compressible and soft materials because of their lower bulk modulus values. Based on the earlier work on solids with van der Waals interactions \cite{Bucko} and also from the present study, we can infer that the peculiar characteristics of these materials are soft and highly compressible.
\subsection{Elastic properties}
Elastic constants are the most important entities in describing the mechanical response of materials. In particular, they are of great importance to understand the response of energetic materials towards the external stimuli such as mechanical shock and thereby their initial step of decomposition or detonation phenomena. Moreover, the knowledge of elastic constants is essential in understanding the high pressure behavior of the materials. Since the elastic properties are strongly related to the lattice parameters optimization results, we use the PBE+G06 functional for the calculation of the elastic constants and compare the results with usual PBE calculations.
Since the metal azides crystallizes in tetragonal symmetry, only six elastic constants are required to describe their mechanical response. The calculated six independent elastic constants of the two metal azides using PBE and PBE+G06 functionals are tabulated in Table 2. The calculated elastic constants of the azides are positive and follow the Born's stability criteria \cite{Born} given by C$_{11}$ $>$ 0, C$_{33}$ $>$ 0, C$_{44}$ $>$ 0, C$_{66}$ $>$ 0, (C$_{11}$-C$_{12}$) $>$ 0, (C$_{11}$ + C$_{33}$- 2C$_{13}$) $>$ 0 and [2(C$_{11}$ + C$_{12}$) + C$_{33}$ +4C$_{13}$] $>$ 0 indicates that the two metal azides are mechanically stable systems. The present results are in good agreement with those calculated from the rigid ion model \cite{Piotr} as well as the ultrasonic measurements on KN$_3$ \cite{Haus}. The computed elastic constant C$_{11}$ of KN$_3$ using PBE+G06 is close to the experimental counterpart \cite{Haus} compared to PBE, which once again confirms the fact of dominance of vdW interactions along a-axis of the lattice over the other axis. In the case of RbN$_3$, there is no experimental data available to compare the present results, but from the optimized crystal structure data we came to a conclusion that the same is true even for RbN$_3$.
 From the calculated elastic constants one can clearly notice that C$_{11}$ is the largest among all the other elastic constants of both KN$_3$ and RbN$_3$. This reflects the fact that the presence of stronger inter molecular interactions along the a-direction or the presence of lowest number of interactions along the c-axis. This suggests that a cleavage plane along c-direction may exist in decomposition of these azides. Moreover, from the calculated elastic constants of KN$_3$ and RbN$_3$, one can also notice that the magnitude of the elastic constants follows the order of RbN$_3$ $<$ KN$_3$, which implies that KN$_3$ is mechanically stiffer system compared to RbN$_3$. The possible reason for this is as follows: the calculated minimum non-bonded distances between M-M, M-N (M=K, Rb) and N-N using PBE and PBE+G06 are presented in Table 3. From the calculated values, one can find that the distances between the atoms increase from KN$_3$ to RbN$_3$, which implies that the interactions among the atoms are weakened as the size of the metal atom increases, hence the stiffness decreases from KN$_3$ to RbN$_3$.
 Linear compressibility is an important mechanical property of the system which gives information about the measure of response of lattice towards hydrostatic stress. The knowledge of elastic constants can be used to calculate the linear compressibilities $\chi_a$ and $\chi_c$ along a, c-axes and volume compressibility $\chi$ via through the empirical relations given in ref \cite{Neumann}. The calculated values of linear compressibilities within PBE and PBE+G06 functionals are tabulated in Table 4.
Since $\chi_c$ value is higher than $\chi_a$, the lattice of the azides would be more compressible along the [0 0 1] direction.
\paragraph*{} As a matter of fact, it is worth to describe the mechanical response of polycrystalline materials. In the present case we calculated the macroscopic mechanical properties of the azides via Voigt-Reuss-Hill approach \cite{Hill} and are tabulated in Table 4.
For any system the association of a high bulk modulus value leads to the lower hydrostatic compressibility. From the calculated bulk modulus value of the two azides it is found that KN$_3$ has higher bulk modulus value implying that the system has lowest compressibility and harder than RbN$_3$.  However, the trend in the bulk modulus value of the azides can be attributed due to the fact that as the ionic radius of the metal atom increases from K to Rb the repulsive interaction between the N-N decreases, see Table 3. Hence RbN$_3$ has lower bulk modulus value compared to KN$_3$. It should be noted here that besides the metal azides with vdW interactions there are also some covalent compounds with high compressibility \cite{Vegas}. AgClO$_4$, a known covalent material has high compressibility because of the presence of Ag atom which weakens the Ag-O bond \cite{Vegas}. Therefore based on the compressibilities of KN$_3$, RbN$_3$ and AgClO$_4$, it can be noticed that the size of the metal cation would also favor for high compressibilities besides the type of bonding present in the compounds. In general the hardness of a material depends on the resistance of the material towards the penetration of another body through the surface. Therefore an important factor of hardness is the strength of the interatomic bonds with respect to the shear deformation, which influences the mobility of dislocations in the solids. In the present case the shear modulus of the metal azides increases in the following sequence G$_{H}$(RbN$_3$) $<$ G$_{H}$(KN$_3$).
\paragraph*{} To know the ductile-brittle nature of the metal azides KN$_3$ and KN$_3$ we have used Pugh's criterion \cite{Pugh}, according to which the critical value of B/G ratio that separates the ductile and brittle material is 1.75. For the present metal azides this value is calculated to be 1.67 for KN$_3$ and 1.74 for RbN$_3$ respectively. These values are close to the vicinity of 1.75 which means that the brittleness of the azides decreases from K to Rb azides, suggesting that KN$_3$ is more brittle than RbN$_3$.
Poisson's ratio is another quantity which is associated with the volume change during the uniaxial deformation.
The values of the Poisson's ratio $\sigma$ for covalent materials are small ($\sigma$ = 0.1), whereas for ionic materials a typical value of $\sigma$ is 0.25. In the present case, the value of $\sigma$ calculated within PBE and PBE+G06 functionals for KN$_3$ is about 0.25 (0.26) and it is 0.26 (0.27) for RbN$_3$ i.e., a considerable ionic contribution in intra-atomic bonding can be assumed.
Young's modulus (E) provides a measure of the stiffness of the solid and if the magnitude of E is large, then the material can be regarded as stiffer material. The calculated E of the two azides is relatively small when compared to the alkali halides whose Young's modulus is of the order of 39 GPa \cite{Bridgman} and therefore the two azides are considered to have lower stiffness than alkali halides. Nevertheless, KN$_3$ is the stiffest material than RbN$_3$ due to its higher E value.
\paragraph*{}
In addition, we also made an attempt to calculate the sound velocities and Debye temperature of the metal azides using the expressions given in ref \cite{VKanchana}. The calculated sound velocities and the Debye temperature are tabulated in Table 5. Our calculated Debye temperature value of KN$_3$ using PBE+G06 functional ($\theta$$_D$ = 361.86 K) is in excellent agreement with the experimental value of $\theta$$_D$ = 351 K \cite{ZIqbal}. This once again prove that the present van der Waals corrected density functional calculations are more precise and reliable. The calculated Debye temperature of the metal azides decreases in the following order: $\theta$$_D$(KN$_3$)$>$ $\theta$$_D$(RbN$_3$). The larger sound velocities and Debye temperature of KN$_3$ leads to high thermal conductivity over RbN$_3$.
\paragraph*{} Recently Haycraft et al \cite{Haycraft} made a correlation of elastic constants of secondary explosives RDX and HMX with their detonation. Their study reveals that RDX has larger elastic constants over HMX implying that RDX is more stiffer system compared to HMX and therefore concluded that HMX is more sensitive towards mechanical shock than that of RDX. This is in very good agreement with the fact that HMX is more sensitive towards shock initiated detonation than RDX. In the case of PETN and RDX, the elastic constants of PETN crystal \cite{Winey} are found to be much smaller than those of RDX and PETN is found to be more sensitive than RDX towards shock initiation. Moreover, among the elastic constants of PETN, C$_{11}$ is found to be the stiffest elastic constant (larger in magnitude) and thereby the detonation is less sensitive along  the [1 0 0] direction \cite{Yoo}. By taking the reference of these studies on correlation of elastic constants with the detonation sensitivity of solid energetic materials, in this present work we have made an attempt to understand the decomposition of the metal azides under present study. From the calculated elastic constants of the metal azides, it was found that the elastic constants of KN$_3$ are larger compared to those of RbN$_3$. They follows the order of RbN$_3$ $<$ KN$_3$, implying that KN$_3$ is mechanically stiffer system than the rubidium azide. This fact also supported from the calculated bulk elastic moduli namely E, B, G of the azides as their magnitude follows the order KN$_3$ $>$ RbN$_3$. In addition, both the azides are found to be stiffer along a-axes followed by c-axes and therefore the lattice can be easily compressed along c-axes over a-axes. This clearly shows that the lattices are most sensitive towards the external stimuli along the [0 0 1] crystallographic direction in these azides. Over all our study based on elastic constants and different elastic moduli reveals that RbN$_3$ is more sensitive towards the mechanical shock initiation over KN$_3$.  
To support this fact we have calculated the decomposition temperature (T$_m$) of the azides by using the elastic constants through the formula given by Fine et al \cite{Fine}. T$_m$ = 354 + 4.5 (2C$_{11}$ + C$_{33}$)/3. The calculated decomposition temperatures are presented in Table 5 along with the experimental values \cite{Mohan, Pistorius}. Our calculated values of T$_m$ using PBE+G06 functional are in fair agreement with that of experimental data and qualitatively reproduce the experimental trend of T$_m$ (RbN$_3$) $<$ T$_m$ (KN$_3$).
\subsection{Electronic properties}
In the case of metal azides it is quite necessary to know about the electronic structure and related properties as they give important information regarding the electronic processes that are responsible for the decomposition and initiation of the metal azides \cite{Evans}. The electronic band structure of KN$_3$ and RbN$_3$ and are shown in Fig. 2 and Fig. 3 respectively. The top of the valence band and the bottom of the conduction band occurs at $\Gamma$-point with a separation of 4.08 eV for KN$_3$ and 4.07 eV for RbN$_3$ which indicates that both the metal azides are direct band gap insulators and the magnitude of the calculated gap agrees well with the earlier theoretical reports \cite{Gord, Zhu}. As the band gap lies in Ultra-Violet (UV) region, the compounds release N$_2$ gas when they irradiated with the UV light. However, we expect that our calculated band gaps might be underestimated from the experimental data, which is a common feature in all density functional theory based calculations \cite{Jones}. The nature of bonded interactions can be well understand by the knowledge of total and partial density of states as they clearly gives the idea of the origin of various bands in terms of atoms and orbitals. The total and partial density of states of the metal azides calculated using PBE+G06 functional is shown in Fig. 4 and Fig. 5 respectively. The valence band spectra and the conduction band spectra of the metal azides show similar features. In the valence band region, there are three regions of bands separated by large gaps which is also evidenced from the energy band structure of the metal azides. There is a strong hybridization between the mid nitrogen atom states and the end nitrogen atom states in the lower energy region of the valence band i.e., from -4 eV to -6.5eV. This indicates that the nitrogen atoms are in strong covalent bonding in the azide ion of the lattice whereas the metal atom states are dominant in the conduction band and there is low overlap of the states with the nitrogen atom states in the valence band, a feature that ionic bonding preserves between the metal atom and the nitrogen atoms. At the Fermi level, the p-states of the end nitrogen atom are only present implying that these states are responsible for decomposition of the azides for an external perturbation.
\paragraph*{} A quantitative picture of the bonded interactions can be obtained from the Mulliken bond population analysis \cite{Mulliken}. The total overlap (bond) population for any pair of atoms in a molecule is in general made up of positive and negative contributions. If the total overlap population between two atoms is positive, they are bonded; if negative, they are anti-bonded. The computed Mulliken bond population of K-N is -0.38 and that of Rb-N is -0.17 whereas the same for N-N is 1.31 and 1.29 of KN$_3$ and RbN$_3$, respectively. The computed Mulliken charges for individual atoms, K (1.11) and N (-0.55) for KN$_3$ and Rb (1.08) and N (-0.53). These results strongly suggest that both the metal azides KN$_3$ and RbN$_3$ are dominantly ionic in nature.
\subsection{Born effective charges and dielectric constants}
The Born effective charge tensor, Z$^*$$_{k, \alpha \beta}$, is a fundamental quantity for the study of lattice dynamics and are defined as the force in the direction $\alpha$ on the atom k due to an homogeneous unitary electric field along the direction $\beta$ or equivalently as the induced polarization of the solid along the direction $\alpha$ by a unit displacement in the direction  $\beta$ of the atomic sub-lattice. Born effective charges also important to characterize the degree of covalency or ionicity in a crystal. By using density functional perturbation theory \cite{Gonze, Refson}, these can be calculated as second derivative of total energy with respect to atomic displacement or to an external field. The calculated Born effective charges are presented in Table 6. From the calculated values we can observe that the Born effective charges of metal atom are close to that of their formal charge +1 and the effective charges are increasing from KN$_3$ to RbN$_3$ implying that the ionicity increases from KN$_3$ to RbN$_3$. The deviation of the calculated Born effective charges of mid N and end N from their formal charges -1 and -2 implies that the charge sharing is dominating over N-N bond. We have also calculated the high and low-frequency dielectric constants of KN$_3$ and RbN$_3$ by the density functional perturbation theory (DFPT) calculations and are presented in Table 7. The calculated high-frequency and static dielectric constants are diagonal and have two independent components as expected for tetragonal symmetry. The electronic components are $\epsilon_{xx}^\infty$ =  $\epsilon_{yy}^\infty$ = 2.74 (2.76) and $\epsilon_{zz}^\infty$ = 2.01 (2.03) for KN$_3$ whereas $\epsilon_{xx}^\infty$ =  $\epsilon_{yy}^\infty$ = 2.68 (2.83) and $\epsilon_{zz}^\infty$ = 2.04 (2.14) for RbN$_3$ within PBE (PBE+G06) respectively. The static components are $\epsilon_{xx}^0$ =  $\epsilon_{yy}^0$ = 5.42 (5.67) and $\epsilon_{zz}^0$ = 5.20 (4.32) for KN$_3$ whereas $\epsilon_{xx}^0$ =  $\epsilon_{yy}^0$ = 6.04 (6.39) and $\epsilon_{zz}^0$ = 4.01 (5.07) for RbN$_3$ within PBE (PBE+G06) respectively. The calculated values of high-frequency and static dielectric constants of KN$_3$ are in good agreement with those of the parameters used for damped oscillator fits to the reflection spectra of KN$_3$ measured in ref \cite{Massa}. There is no experimental data available for RbN$_3$ to compare the present theoretical results. The averages of  $\epsilon^\infty$ and $\epsilon^0$ obtained from the expression $\epsilon^\infty$ ($\epsilon^0$) = (2$\epsilon_{zz}^\infty$ + $\epsilon_{xx}^\infty$)/3. The calculated values of the average of $\epsilon^\infty$ for KN$_3$ and RbN$_3$ are 2.25 (2.27) and 2.25 (2.37)  within PBE (PBE+G06) respectively. The average of $\epsilon^0$ for KN$_3$ and RbN$_3$ within PBE (PBE+G06) are 5.27 (4.77) and 4.68 (5.51), respectively. These results indicate there is a considerable anisotropy which eventually suggests that both the metal azides have large tetragonal distortion.
\subsection{Phonon density of states and Phonon frequencies at $\Gamma$-point}
In this section we will discuss about the calculated phonon density of states and phonon frequencies at gamma point of the metal azides under study. The phonon density of states calculated through the density functional perturbation theory \cite{Gonze, Refson} using the PBE+G06 functional are shown in Fig. 6 and Fig. 7 respectively. The lower frequency region states are dominated by the metal atoms whereas the higher frequency region are entirely due to the azide ion alone, the fact can be attributed to the difference in their individual masses. The important point to be notice from the phonon density of states is that around 100 cm$^{-1}$, the states of metal atom and azide ion are overlapping each other. This implies that the vibrations are due to the collective excitations of both metal and azide ion respectively.
The primitive cell of KN$_3$ and RbN$_3$ consists of eight atoms and hence they have 24 vibrational modes out of which three are acoustic modes and 21 are optical modes. According to the symmetry analysis of the point group D$_{4h}$, the acoustic and optical modes at the $\Gamma$ point can be classified into the following symmetry species:\\
$\Gamma_{aco}$ =  A$_{2u}$ + 2E$_{u}$\\
$\Gamma_{opt}$ = 2A$_{2g}$ + 2A$_{2u}$ + A$_{1g}$ + 4E$_{g}$ + 8E$_{u}$ + B$_{1g}$ + 2B$_{1u}$ + B$_{2g}$\\
  Out of these modes, the A$_{2g}$ vibrations are silent as they do not cause a change in polarizability or dipole moment and therefore these modes are optically inactive. The modes A$_{1g}$, B$_{2g}$, and B$_{1g}$ are Raman active (RA) whereas the A$_{2u}$, B$_{1u}$, and E$_{1u}$ modes are infrared active (IA). The vibrational frequencies calculated at the theoretical equilibrium volume using PBE and PBE+G06 functionals are presented in Table 8 and 9 along with the experimental data. Since the metal azide crystal consists of tightly bounded azide ions which are loosely bounded to metal atoms, the vibrations involving the nitrogen atoms of azide ion could be labeled as internal vibrations and the external or lattice vibrations are those in which the azide ion move as rigid units along with the metal ion sub lattice. The calculated frequencies ranging from 89 cm$^{-1}$ to 200 cm$^{-1}$ of KN$_3$ and 56 cm$^{-1}$ to 179 cm$^{-1}$ of RbN$_3$ are due to lattice mode vibrations, for which the results of PBE+G06 functional are in fair agreement with experiment \cite{Massa, Hath, Bryant, Govinda} than PBE values. This is because the intermolecular van der Waals forces strongly couples in lattice modes which PBE could not take in to account. But, for the bending mode frequencies of internal modes due to the azide ions, we find a very good agreement with experiment \cite{Lam, Iqbal, Papa} using both PBE and PBE+G06 functionals. This might be due to the fact that these vibrations are purely due to the motion of individual N atoms of each azide ion which are covalently bonded with each other. Hence both the functionals give similar frequencies compared to experimental data. However, the stretching mode frequencies of the azide ion ranging from 1200 cm$^{-1}$ to 1900 cm$^{-1}$ are underestimated by 8.4$\%$ for A$_{1g}$ and B$_{1g}$ modes (both are due to the symmetric stretching of azide ion) and the E$_{u}$ mode (asymmetric stretching of azide ion) is underestimated by 8.4$\%$ in both the metal azides. This might be due to the fact that our present calculations are performed through the linear response approach which is solely based on the harmonic approximation and therefore the anharmonicity present in the higher frequencies which are mainly due to the azide ion could not be dealt efficiently with the calculations.
 \subsection {Thermodynamic properties}
Thermodynamic properties are very important, as they play a major role in understanding the thermal response of the solids. According to the standard thermodynamics, if the system is held at a fixed temperature T and pressure P then the Gibb's free energy (G) of the system is expressed as
G(V,P,T) = E(V) + PV-TS
where E(V) is the total energy, V is volume and S is entropy of the system. Since the electronic structure calculations are performed in the static approximation, i.e., at T=0 K and neglecting zero-point vibrational effects, the corresponding Gibb's free energy in this case becomes $G_{stat}(V,P) = E(V)+PV$, which is the enthalpy H of the system. But the experimental determination of thermodynamical properties takes place at finite T and therefore the vibrational effects should be considered. Hence it is necessary to include these effects in order to compare theoretical predictions with the experimental measurements to explore the thermodynamic properties. So we have chosen the quasi-harmonic Debye model \cite{Blanco} and according to this model the non-equilibrium Gibbs free energy function is given by
G$^*$(V,P,T)=E(V)+PV+A$_{vib}$ ($\Theta(V)$, T)
where $\Theta(V)$ is the Debye temperature, PV corresponds to the constant hydrostatic pressure condition and $A_{vib}$ is the vibrational Helmholtz free energy and the corresponding expressions can be found elsewhere in \cite{Blanco}.
The calculated total phonon density of states, as discussed in earlier section, were used to obtain the temperature dependent heat capacity of the metal azides. Usually, the quasi harmonic theory does not valid when the temperature is close to the melting point of the solids \cite{Baroni}. Therefore, we have used the temperature range from 0 to 500 K which is below the melting point of the metal azides, KN$_3$  (618 K) and RbN$_3$ (590 K). The calculated heat capacity of the metal azides is shown in Fig. 8. The heat capacity of RbN$_3$ is quite higher compared to that of KN$_3$, implies the thermal response of the  RbN$_3$ system is more compared to that of KN$_3$ solid. The important thing to be noticed in the calculated heat capacities of both the azides is that they are in good agreement with experiment \cite{Carling} at low temperatures rather than at high temperatures. This is because all the density functional theory calculations are carried out at 0K, hence one would expect a good agreement with experiment at low temperatures.
\section{Conclusions}
In conclusion, we have studied the structural, bonding, elastic and vibrational properties of energetic molecular crystals namely alkali metal azides KN$_3$ and RbN$_3$. We have focussed our attention more towards the effect of van der Waals interactions on the structural properties, elastic stiffness tensor, and vibrational frequencies of the metal azides. We find that all properties are improved to a great extent by the inclusion of vdW interactions in the calculations. In particular, PBE+G06 functional gives good results compared to other dispersion corrected functionals. The calculated elastic constants suggest that the metal azides are mechanically stable systems under normal conditions. The elastic constants of KN$_3$ are larger in magnitude than those of RbN$_3$, implying that potassium azide is mechanically stiffer material over rubidium azide. This fact is also confirmed by the higher values of bulk modulus, shear modulus and Young's modulus of KN$_3$ over RbN$_3$. Moreover, both KN$_3$ and RbN$_3$ are found to be brittle in nature. From the calculated elastic constants we obtain the decomposition temperature (T$_m$) of the azides and found that T$_m$ (RbN$_3$) $<$ T$_m$ (KN$_3$). Overall from the study of elastic constants we can come to a conclusion that RbN$_3$ is more sensitive towards a  mechanical shock as the compound has lower elastic moduli than KN$_3$. The vibrational frequencies, in particular the frequencies of lattice modes are well reproduced by the present van der Waals corrected density functional calculations. Whereas, the higher frequency modes which are entirely due to the vibration of azide ion are well described by both the standard DFT and van der Waals corrected DFT functionals. In addition, we also presented the heat capacity of both the azides calculated through van der Waals corrected DFT, which is in excellent agreement with experiment at lower temperatures.
\section{ACKNOWLEDGMENTS}
 K R B would like to thank DRDO through ACRHEM for financial support. The authors acknowledge CMSD, University of Hyderabad for providing computational facilities.

\newpage
Figure Legends\\
Figure 1: (Colour online) Layered structure of metal azide MN$_3$ (M= K or Rb). In figure, violet ball indicates, metal atom and blue ball indicates nitrogen atom, respectively.\vspace{1cm} \\
Figure 2:(Colour online) Energy band structure of KN$_3$ calculated with PBE+G06 functional \vspace{1cm}\\
Figure 3:(Colour online) Energy band structure of RbN$_3$ calculated with PBE+G06 functional \vspace{1cm}\\
Figure 4: (Colour online) Density of states of KN$_3$ calculated with PBE+G06 functional \vspace{1cm}\\
Figure 5: (Colour online) Density of states of RbN$_3$ calculated with PBE+G06 functional \vspace{1cm}\\
Figure 6: Total and Partial phonon density of states of KN$_3$ calculated within PBE+G06 functional at theoretical equilibrium volume \vspace{1cm}\\
Figure 7: Total and Partial phonon density of states of RbN$_3$ calculated within PBE+G06 functional at theoretical equilibrium volume \vspace{1cm}\\
Figure 8: (Colour online) Heat capacity of metal azides MN$_3$ (M=K, Rb) in cal/mol.K calculated within PBE+G06 functional at theoretical equilibrium volume
\clearpage
\newpage
\begin{figure}
\centering
\caption{(Colour online) Layered structure of metal azide MN$_3$ (M= K or Rb). In figure, violet ball indicates, metal atom and blue ball indicates nitrogen atom, respectively.}\label{Fig 1}
\begin{tabular}{cc}
\epsfig{file=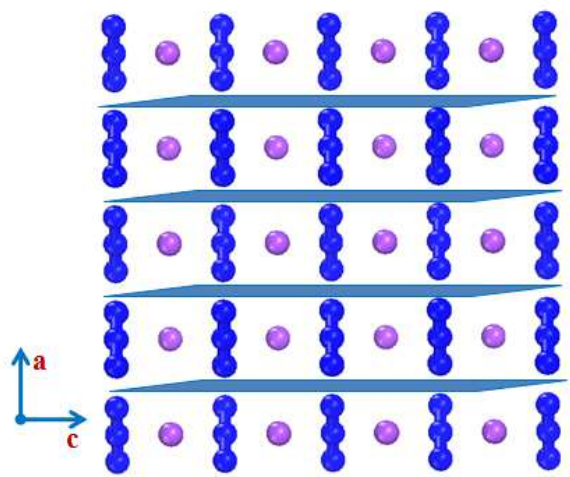,width=0.65\linewidth,clip=} &\\
\end{tabular}
\end{figure}
\clearpage

\begin{figure}
\centering
\includegraphics[width=15cm]{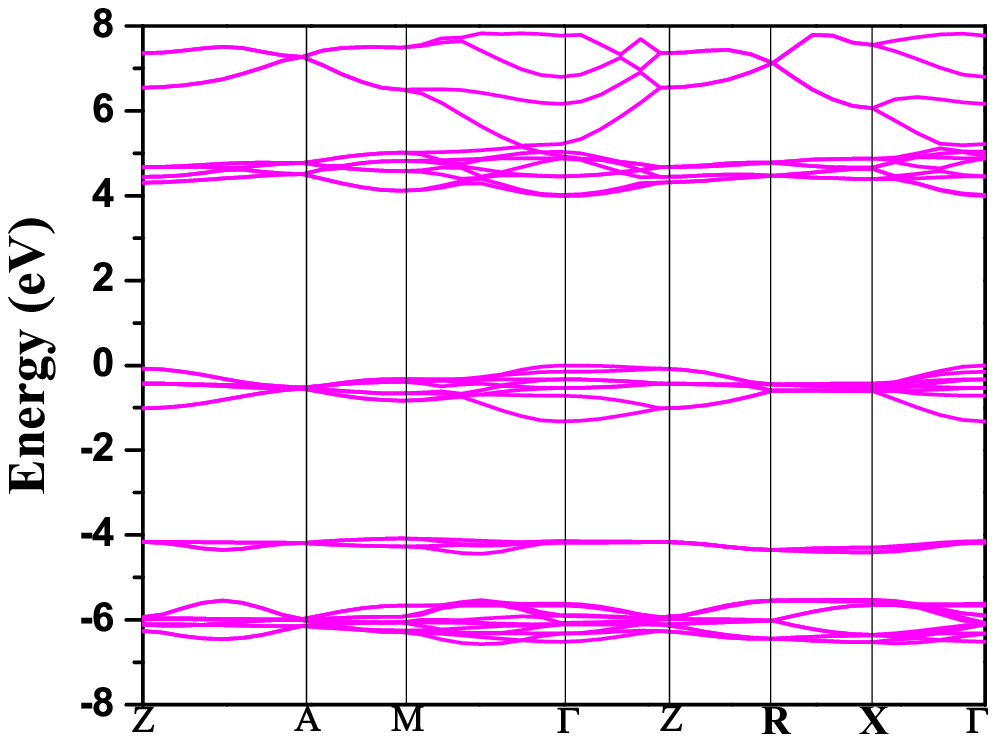}\\
 \caption{(Colour online) Energy band structure of KN$_3$ calculated with PBE+G06 functional}
 \end{figure}
 \clearpage

 \begin{figure}
\centering
\includegraphics[width=15cm]{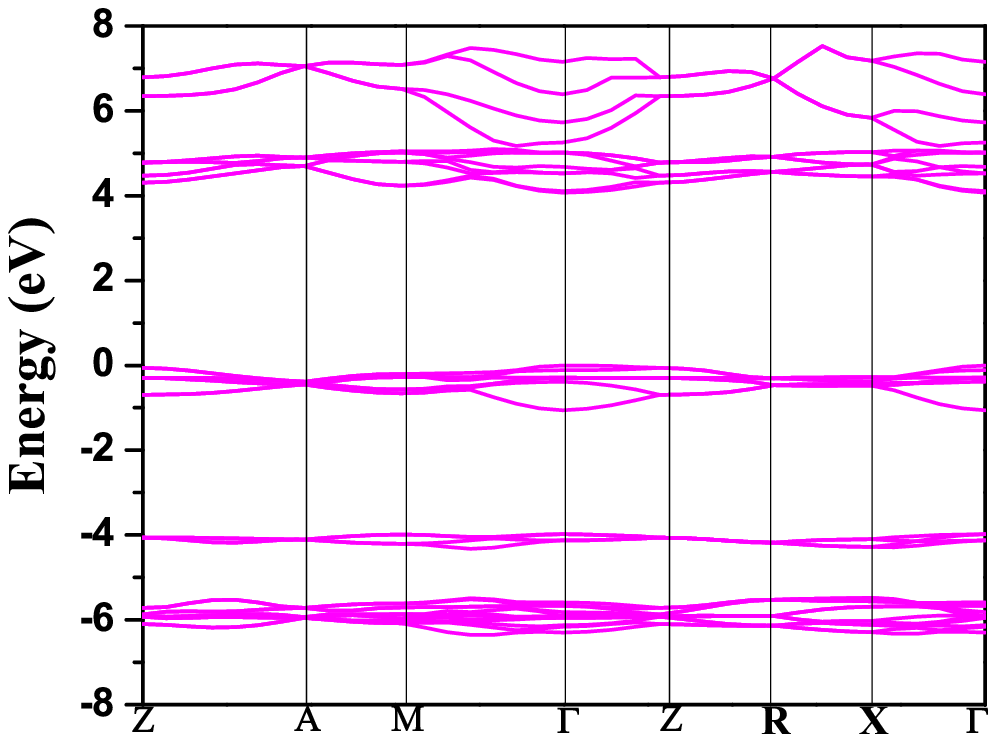}\\
 \caption{(Colour online)  Energy band structure of RbN$_3$ calculated with PBE+G06 functional}
 \end{figure}
 \clearpage

\begin{figure}
\centering
\includegraphics[width=15cm]{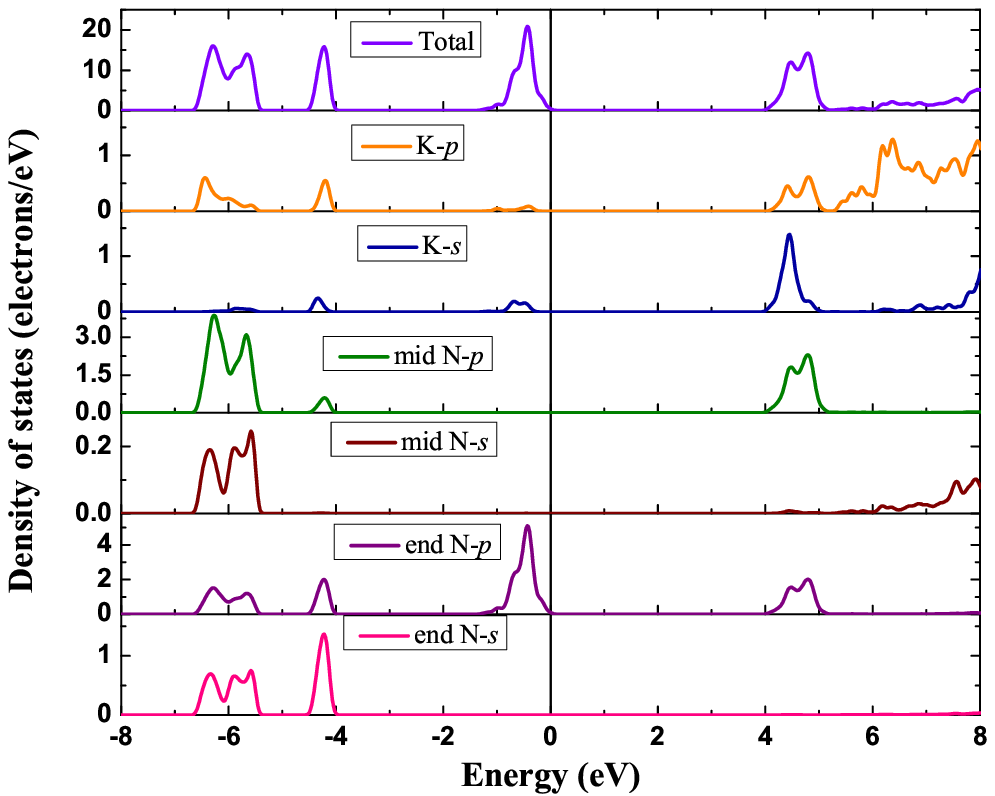}\\
 \caption{(Colour online) Density of states of KN$_3$ calculated with PBE+G06 functional}
 \end{figure}
 \clearpage

 \begin{figure}
\centering
\includegraphics[width=15cm]{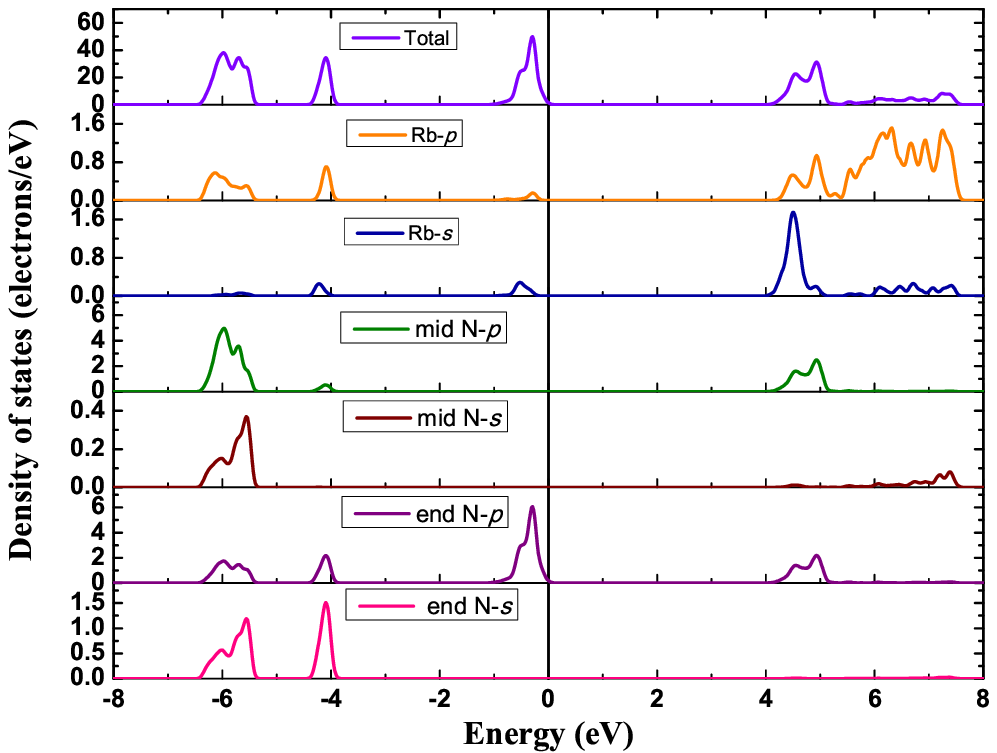}\\
 \caption{(Colour online) Density of states of RbN$_3$ calculated with PBE+G06 functional}
 \end{figure}
 \clearpage


 \clearpage
 \begin{figure}
\centering
\includegraphics[width=15cm]{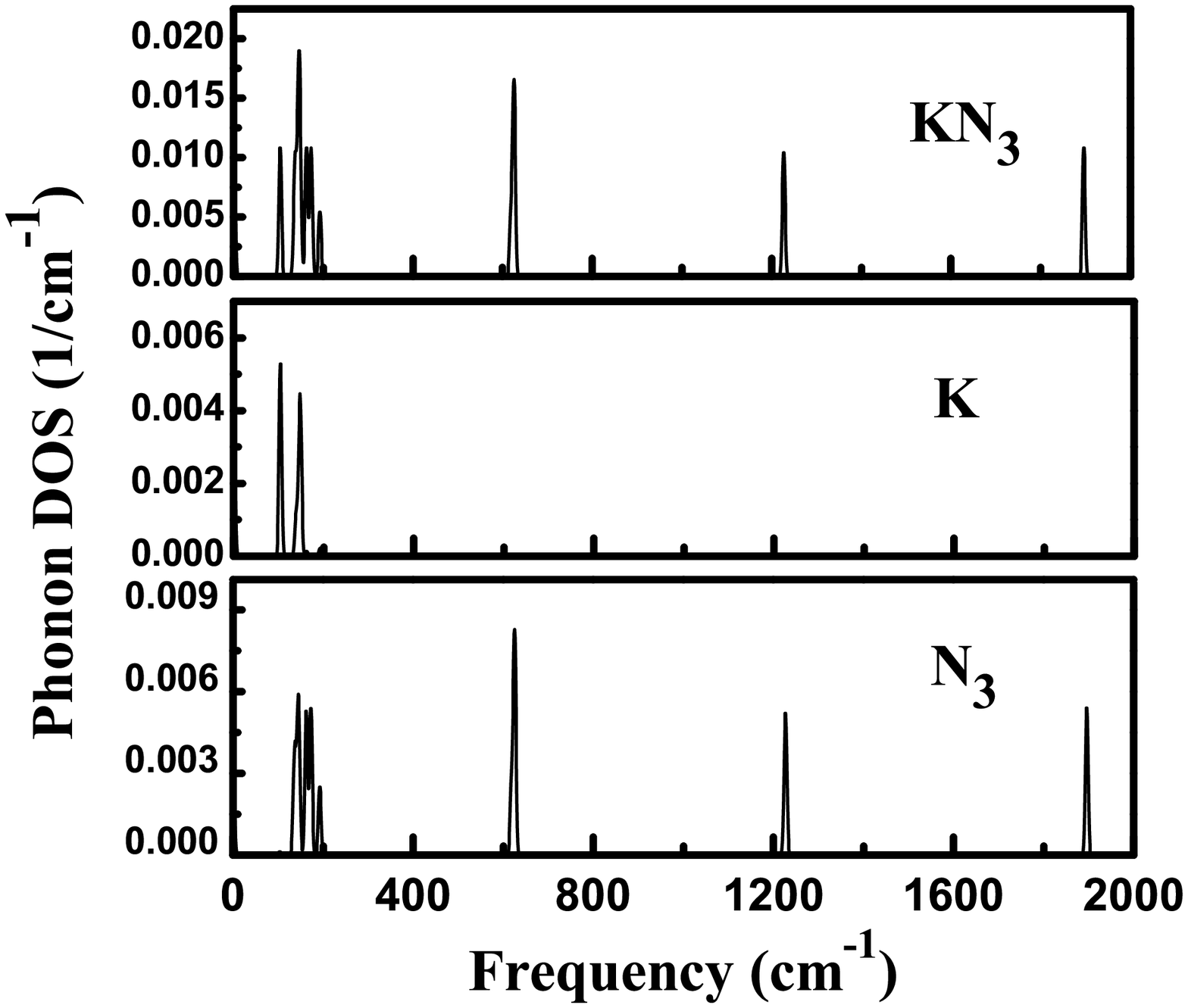}\\
 \caption{ Total and partial phonon density of states of KN$_3$ calculated within PBE+G06 functional at theoretical equilibrium volume.}\label{Fig 1}
 \end{figure}

 \clearpage
 \begin{figure}
\centering
\includegraphics[width=15cm]{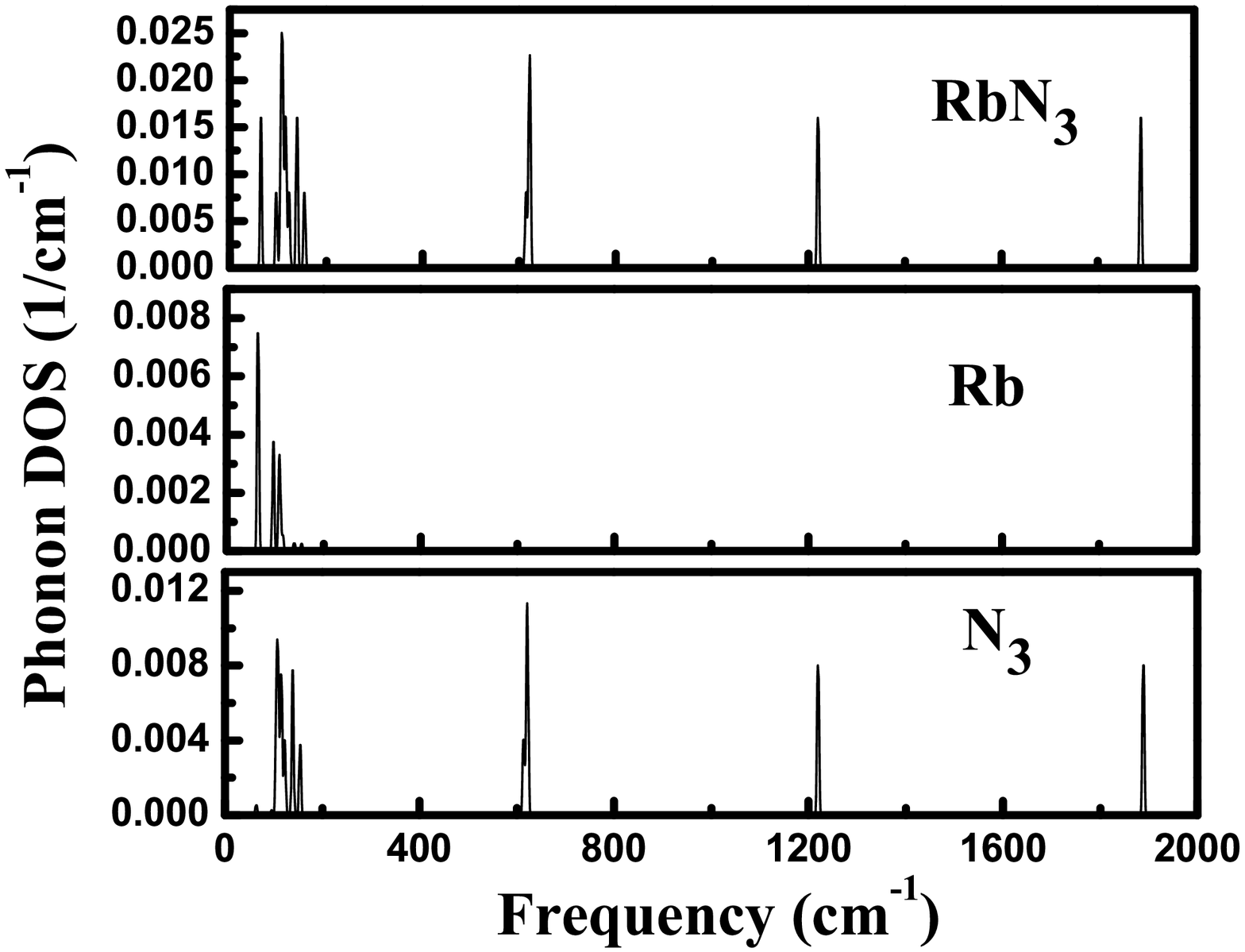}\\
 \caption{ Total and partial phonon density of states of RbN$_3$ calculated within PBE+G06 functional at theoretical equilibrium volume.}
 \end{figure}
 \clearpage
 \begin{figure}
\centering
\includegraphics[width=15cm]{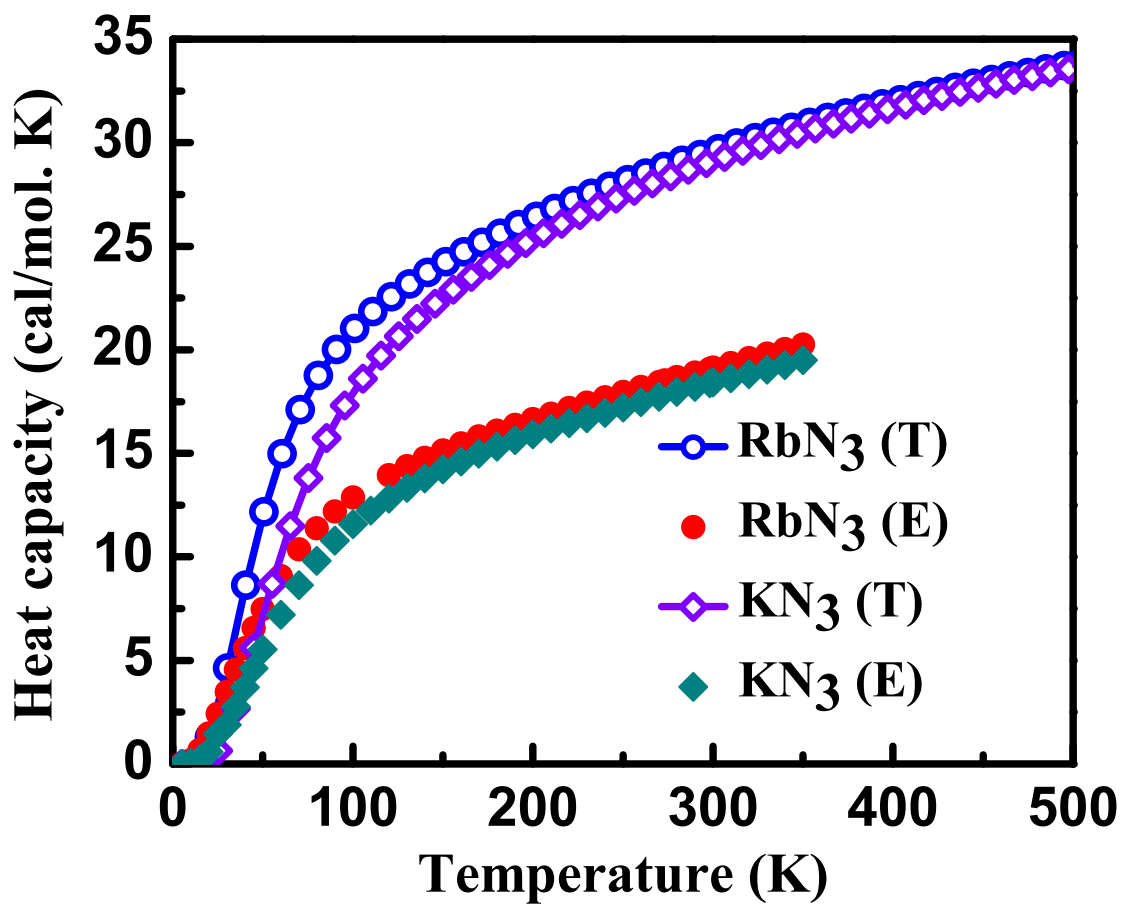}\\
 \caption{(Colour online) Heat capacity of metal azides MN$_3$ (M=K, Rb) in cal/mol.K calculated within PBE+G06 functional at theoretical equilibrium volume.}
 \end{figure}
\newpage
\clearpage
\newpage
\begin{table}
\caption{The ground state properties of tetragonal KN$_3$, RbN$_3$ calculated using LDA (CA-PZ), GGA (PBE), LDA (OBS), GGA (PBE+TS) and GGA (PBE+G06) functionals. Experimental data are taken from Ref \cite{Muller}, except where otherwise stated.}
\label{tab.1}
\begin{tabular}{ccccccc}\hline
Method& a(\AA)&c(\AA) & V (\AA)&B$_0$ (GPa) &$\beta$ (GPa$^{-1}$)\\ \hline
KN$_3$&  & & & & &\\
LDA (CA-PZ)&5.898&6.653  &231.38&30.3&0.033\\
GGA (PBE)&6.205  &7.207  &277.54&18.4&0.054\\
LDA (OBS) & 5.817 & 6.430 & 217.6 & 36.7 & 0.027 \\
GGA (PBE+TS)& 5.620 & 7.309 & 230.8 & 52.1 & 0.019 \\
GGA (PBE+G06)&6.102&6.961&259.16&27.2&0.036\\
Expt&6.113&7.094&265.09& 27.4$^a$&--\\ \hline
RbN$_3$&  & & & & &\\
LDA(CA-PZ)&6.115&7.161 &267.75&27.3&0.036\\
GGA(PBE)&6.445  &7.691  &319.54&15.8&0.063\\
LDA (OBS) & 6.042 & 6.979 & 254.8 & 32.4 & 0.030\\
GGA (PBE+TS) & 5.761 & 7.541 & 250.3 & 37.5 & 0.026\\
GGA(PBE+G06)&6.359&7.407&299.49&25.2&0.039\\
Expt&6.310&7.519&299.37&--&--\\ \hline
$^a$Ref \cite{Medvedev-2}
\end{tabular}

\end{table}
\clearpage
\newpage
\begin{table}
\caption{Second-order elastic constants, C$_{ij}$ in GPa, of MN$_3$ (M=K, Rb) calculated within GGA (PBE) and GGA (PBE+G06) functionals.}
\begin{tabular}{ccccccc}\hline \hline
Method& C$_{11}$  & C$_{12}$ & C$_{13}$ & C$_{33}$ & C$_{44}$ & C$_{66}$ \\ \hline
& & KN$_3$& & & &\\ \hline
GGA(PBE)&39.4&15.3&7.5&31.7&7.3&18.6 \\
GGA(PBE+G06)&54.5&23.1&16.4&37.3&15.3&20.2\\
Expt$^p$ &49.5&13.9&11.7&32.3&8.8&13.1 \\ \hline
& & RbN$_3$& & &  & \\ \hline
GGA(PBE)&30.4&12.4&7.7&27.4&6.8&13.1\\
GGA(PBE+G06)&43.3&17.8&17.6&36.3&15.1&13.1\\ \hline
$^p$Ref \cite{Haus}
\end{tabular}
\end{table}
\clearpage
\newpage
\begin{table}
\caption{The minimum non bonded distance of M-M, M-N, and N-N (M=K and Rb) of tetragonal KN$_3$ and RbN$_3$ calculated using GGA-PBE and GGA (PBE+G06) functionals.}
\label{tab.1}
\begin{tabular}{cccc}\hline \hline
Bond & KN$_3$  & RbN$_3$ \\ \hline
$M-M$ (\AA) & 3.604 (3.481) & 3.846 (3.706) \\
$M-N$(\AA) & 3.002 (2.982) & 3.164 (3.078)  \\
$N-N$(\AA) & 3.169 (3.096) & 3.338 (3.255) \\ \hline
\end{tabular}
\end{table}
\clearpage
\newpage
\begin{table}
\caption{Polycrystalline properties of MN$_3$ (M=K, Rb) B, G, E in GPa and linear compressibilities, $\chi$$_{a}$, $\chi$$_{c}$ and $\chi$ in (TPa)$^{-1}$, calculated within GGA (PBE) and GGA (PBE+G06) functionals.}
\begin{tabular}{ccccccccccc}\hline \hline
Method&B$_{V}$  & B$_{R}$ & B$_{H}$ & G$_{V}$ & G$_{R}$ & G$_{H}$& E&$\chi$$_{a}$ &$\chi$$_{c}$ &$\chi$ \\ \hline
& & KN$_3$& & &  & &&&&\\ \hline
GGA(PBE)&19.0&18.4&18.7&11.9&10.4&11.2& 28.0&14.9&24.5&54.3\\
GGA(PBE+G06)&28.6&27.2&27.9&16.2&15.7&15.9& 40.1&8.8&19.1&36.7\\ \hline
& & RbN$_3$& & &  & &&&&\\ \hline
GGA(PBE)&15.9&15.8&15.9&9.4&8.7&9.1&22.9&18.6&25.9&63.1\\
GGA(PBE+G06)&25.4&25.2&25.3&13.3&13.1&13.2&33.7&11.6&16.2&39.4\\ \hline
\end{tabular}
\end{table}
\clearpage
\newpage
\begin{table}
\caption{The longitudinal ($\upsilon$$_{l}$), transverse ($\upsilon$$_{t}$) and average ($\upsilon$$_{m}$) sound velocities, Debye temperature ($\theta$$_D$) and decomposition temperature (T$_m$) of tetragonal KN$_3$, RbN$_3$ calculated within GGA-PBE and GGA (PBE+G06) functionals.}
\label{tab.1}
\begin{tabular}{ccc}\hline \hline
Property & KN$_3$  & RbN$_3$ \\ \hline
$\upsilon$$_{l}$ (km/s) & 4.16 (4.86) & 3.25 (3.88) \\
$\upsilon$$_{t}$(km/s) & 2.39 (2.77)& 1.84 (2.15) \\
$\upsilon$$_{m}$(km/s) & 2.66 (3.07)& 2.05 (2.39)\\
$\theta$$_{D}$ (K) & 306.16 (361.86) [351]$^c$ & 225.37 (269.79)\\
T$_m$ (K) & 519 (574) [618]$^d$ & 486 (539)[590]$^e$\\ \hline
$^c$Ref \cite{ZIqbal}, $^d$Ref \cite{Mohan}, $^e$Ref \cite{Pistorius}
\end{tabular}
\end{table}
\clearpage
\newpage
\begin{table}
\caption{The Born-effective charges of tetragonal KN$_3 $ and RbN$_3$ calculated within GGA-PBE (GGA-PBE+G06) functionals.}
\label{tab.1}
\begin{center}
\begin{tabular}{ccc}\hline \hline
 & KN$_3$ & RbN$_3$ \\ \hline
Z$^*_{xx}$ &     K: 1.209 (1.241) & Rb: 1.255 (1.301) \\
           & mid N: 1.692 (1.705) & mid N: 1.741 (1.768) \\
           & end N: -1.450 (-1.472) & end N: -1.498 (-1.534)\\
           & & \\
Z$^*_{yy}$ &  K: 1.209 (1.241) & Rb: 1.255 (1.301) \\
           & mid N: 1.692 (1.705) & mid N: 1.741 (1.768) \\
           & end N: -1.450 (-1.472) & end N: -1.498 (-1.534)\\
           & & \\
Z$^*_{zz}$ & K: 1.120 (1.115) & Rb: 1.205 (1.206) \\
           & mid N: -0.130 (-0.122) & mid N: -0.169 (-0.162)\\
           & end N: -0.495 (-0.496) & end N: -0.518 (-0.522)\\
           & & \\
Z$^*_{xy}$ & K: 0 (0) & Rb: 0 (0) \\
           & mid N: 1.759 (1.793) & mid N:1.759 (1.793)\\
           & end N: -0.869 (-0.887) & end N: -0.869 (-0.887)\\
           & & \\
Z$^*_{xz}$ & K: 0 (0) & Rb: 0 (0) \\
           & mid N: 1.759 (1.793) & mid N:1.759 (1.793)\\
           & end N: -0.869 (-0.887) & end N: -0.869 (-0.887)\\
           & & \\   \hline \hline
\end{tabular}
\end{center}
\end{table}
\clearpage
\newpage
\begin{table}
\caption{The dielectric constants of tetragonal KN$_3$, RbN$_3$ calculated within GGA-PBE (PBE+G06) functionals.}
\label{tab.1}
\begin{center}
\begin{tabular}{ccccc}\hline \hline
Compound & &$\epsilon_{xx}$ & $\epsilon_{yy}$ & $\epsilon_{zz}$ \\ \hline
KN$_3$& $\epsilon^\infty$ &2.74 (2.76) [2.75]$^f$&2.74 (2.76) [2.75] $^f$&2.01 (2.03) [3.50]$^f$\\
      & $\epsilon^0$&5.42 (5.67) [6.05]$^f$&5.42 (5.67) [6.05]$^f$&5.20 (4.32) [7.21]$^f$\\
RbN$_3$&  $\epsilon^\infty$&2.68 (2.83)&2.68 (2.83)&2.04 (2.14)\\
      & $\epsilon^0$&6.04 (6.39)&6.04 (6.39)&4.01 (5.07)\\ \hline
$^f$Ref \cite{Massa}
\end{tabular}
\end{center}
\end{table}
\begin{table}
\caption{The calculated vibrational frequencies (cm$^{-1}$) of tetragonal KN$_3$ at ambient pressure within PBE and PBE+G06 functionals. T and R corresponds to Translational and Rotational vibrations respectively. IA and RA infers the Infrared active and Raman active modes of KN$_3$.}
\label{tab.2}
\begin{center}
\begin{tabular} {c c c c}\hline
Mode symmetry& &Frequency (cm$^{-1}$)& Expt \\\hline
& PBE & PBE+G06&  \\
E$_g$(T) (RA) &89.8&103.8&103$^1$\\
B$_{1u}(T)$&104.9&136.5& \\
B$_{1g}$(R)  (RA)&106.1&139.2& \\
A$_{2u}$(T)  (IA)&109.6&144.1&138.2$^2$ \\
E$_{u}$(T)  (IA)&114.7&146.2& \\
E$_{g}$(R)  (RA)&131.5&149.4&145$^1$\\
&  &  &147$^3$\\
A$_{2g}$(T)&134.9&162.8& \\
E$_{u}$(T)  (IA)&155.5&173.4&168$^2$\\
A$_{2g}$(R)&162.3&192.2&  \\
A$_{2u}$  (IA)&617.2&618.2&627$^4$\\
&  &  & 643$^5$ \\
 &  &  &624$^6$ \\
B$_{1u}$&623.1&625.3& \\
E$_{u}$&623.6&626.3& 624$^4$ \\
&  &  &  647$^5$ \\
&  &  &  646$^6$ \\
B$_{2g}$  (RA)&1219.1&1226.3&1339$^1$\\
A$_{1g}$  (RA)&1220.7&1228.5& 1340$^1$\\
E$_{u}$ (IA)&1883.7&1896.1&2002.2$^6$\\ \hline
$^1$Ref\cite{Hath}, $^2$Ref\cite{Massa}, $^3$Ref\cite{Bryant}, $^4$Ref\cite{Lam}, $^5$Ref\cite{Iqbal}, $^6$Ref\cite{Papa}
\end{tabular}
\end{center}
\end{table}
\clearpage
\newpage
\begin{table}
\caption{The calculated vibrational frequencies (cm$^{-1}$) of tetragonal RbN$_3$ at ambient pressure within PBE and PBE+G06 functionals. T and R corresponds to Translational and Rotational vibrations respectively. IA and RA infers the Infrared active and Raman active modes of RbN$_3$.}
\label{tab.2}
\begin{center}
\begin{tabular} { c  c  c c }\hline
Mode symmetry& &Frequency (cm$^{-1}$)& Expt \\\hline
& PBE & PBE+G06& \\
E$_g$(T)  (RA)&56.2&64.8&66$^x$\\
B$_{1u}(T)$&85.5&96.3& \\
B$_{1g}$(R)  (RA)&104.8&105.9& \\
A$_{2u}$(T)  (IA)&119.2&107.6&108$^y$ \\
E$_{u}$(T)  (IA)&131.5&109.3& \\
E$_{g}$(R)  (RA)&149.3&115.3&157$^x$\\
A$_{2g}$(T)&153.8&122.6& \\
E$_{u}$(T)  (IA)&155.2&139.2&168$^y$\\
A$_{2g}$(R)&179.1&155.1&  \\
A$_{2u}$  (IA)&615.7&614.6& 624.5$^z$\\
B$_{1u}$&618.9&619.9& \\
E$_{u}$&621.4&621.8& 643.5$^z$\\
B$_{2g}$  (RA)&1214.8&1219.1&--\\
A$_{1g}$  (RA)&1215.8&1220.3& 1333$^x$\\
E$_{u}$ (IA)&1877.9&1888.7&2008$^z$\\ \hline
$^x$Ref\cite{Hath}, $^y$Ref\cite{Massa}, $^z$Ref\cite{Papa}
\end{tabular}
\end{center}
\end{table}
\clearpage
\end{document}